\documentclass[aps,twocolumn,nopacs,floats,superscriptaddress,widetext]{revtex4-2}
\usepackage{float}
\usepackage{subfigure}
\usepackage{graphicx}
\usepackage{placeins}
\usepackage{dcolumn}
\usepackage{bm}
\usepackage{color}
\def\he4{$^4$He}
\def\hel3{$^3$He}
\def\Am3{\AA$^{-3}$}
\def\beq{\begin{equation}}
\def\eeq{\end{equation}}
\begin{document}


\author{Nikolay V. Prokof'ev}
\affiliation{Department of Physics, University of Massachusetts, Amherst, MA 01003, USA}

\author{Boris V. Svistunov}
\affiliation{Department of Physics, University of Massachusetts, Amherst, MA 01003, USA}
\affiliation{Wilczek Quantum Center, School of Physics and Astronomy and T. D. Lee Institute, Shanghai Jiao Tong University, Shanghai 200240, China}

\title{Phonon-modulated-hopping Polarons: $X$-representation Technique}
\begin{abstract}
Motivated by the  problem of polaron effect due to phonon-modulated hopping,
we formulate a generic Monte Carlo technique for solving it in the coordinate representation
for both the particle and atomic displacements. The method applies to a broad class of models;
the only condition is that the hopping amplitude be sign-positive. A dramatic simplification
of the scheme, with the corresponding efficiency gain, takes place in models with dispersionless phonons.
Our study sheds important light on the nature and universality of the most striking qualitative and quantitative effects demonstrated by the ``standard" (Peierls/Su-Schrieffer-Heeger) model based on the linearized displacement-modulated hopping.
\end{abstract}

\maketitle

\begin{figure}[tbh]
\includegraphics[width=0.45\textwidth]{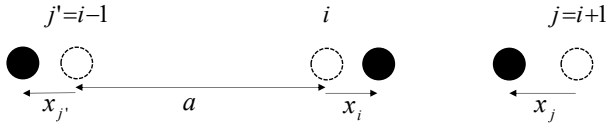}
\caption{In the PSSH model, the tunneling amplitude depends on the distance between the atoms and
is enhanced (suppressed) when atoms move closer to (away from) each other. This leads to the sign-alternating
dependence on the atomic displacement conditional to whether the tunneling takes place in the positive or negative lattice direction, see Eqs.~(\ref{txx1})--(\ref{txx2}). }
\label{Fig1}
\end{figure}

{\it Introduction}. The effect of phonons on the tunneling motion of particles between the sites of an elastic lattice has been of significant interest for a long time \cite{Barisic1970,Barisic1972,KK,SSH,Mona2010,Sous2018,Xing2021,Cai2021,us2021,Sous2021,Gotz2022,Feng2022,Zhang2022a,Zhang2022b}.
Recently, this interest has been boosted by intriguing results for polarons in the Peierls/Su-Schrieffer-Heeger (PSSH) model
based on the displacement-modulated hopping with linear coupling to atomic displacements \cite{Mona2010,Sous2018,us2021,Sous2021},
see an illustration in Fig.~\ref{Fig1}. It was found that PSSH polarons differ dramatically from their conventional counterparts
originating from phonon coupling to the particle density. One significant difference---crucially important in the context of the bipolaron mechanism of high-temperature superconductivity---is the absence of sharp self-trapping crossover and dramatic increase of the effective mass in the adiabatic regime when the phonon frequency is much smaller than the particle bandwidth. Another striking effect demonstrated by PSSH polarons is the shift of the ground-state momentum to finite values when the coupling exceeds a certain critical value \cite{Mona2010,Sous2018,us2021,Sous2021}.

The most general form of the phonon-modulated-hopping term is
\begin{equation}
- t_{ij} \left( \{ x_s \} \right) a_{j}^{\dagger} a_{i}^{\,} ,
\label{hop}
\end{equation}
where $a_{j}^{\dagger}$ and $a_{i}^{\,} $ are the particle creation and annihilators operators on the sites $j$ and $i$, respectively, 
and $t_{ij}$ is the hopping amplitude that depends on a certain set of atomic displacements $x_s$, which, in their turn,
are linear combinations of the phonon creation and annihilation operators. To define the class of models solvable by Monte Carlo (MC) methods we consider the hopping term (\ref{hop}) in the $x$-representation, when the displacement $x_s$, and, thus, the amplitude $t_{ij}$, are real numbers.
In the vast majority of cases---the single-orbital tunneling being a characteristic example---the function
$t_{ij} \left( \{ x_s \} \right) $ is sign-definite. However, when $t_{ij} \left( \{ x_s \} \right)$ is approximated by a
linear function of $\{ x_s \}$, as is typically done for PSSH-type models, the sign-definiteness condition is violated
for large atomic displacements.
This raises two related questions: what is known about PSSH polarons with highly non-linear coupling to phonons,
and whether the violation of the sign-definiteness condition is a key ingredient behind the special  properties
of ``linearized" PSSH polarons.

In this Letter, we provide first-principle numeric evidence that unusual properties characteristic of linearized
PSSH polarons are absent in a non-linear model with sign-positive $t_{ij}$. Our MC approach is based on the observation that  sign-definiteness of $t_{ij}$ allows one to employ a sign-positive path integral formulation in the $x$-representation.
We study the standard case of dispersionless phonons here when substantial part of the path integral
is performed analytically resulting in a simple and efficient diagrammatic-type ground-state MC scheme.

{\it Model Hamiltonian.} The above-mentioned special properties of PSSH polarons are captured by the linearized
one-dimensional PSSH model with dispersionless phonons \cite{Mona2010,Sous2018,us2021,Sous2021}. This motivates
us to consider a positive-$t_{ij}$ extension of the PSSH model:
\begin{equation}
H=H_1+H_2
=-\sum_{<ij>} t(x_i,x_j)
a_{j}^{\dagger} a_{i}^{\,}
+ \Omega \sum_i b_i^{\dagger} b_i^{\,} \, ,
\label{H}
\end{equation}
were $b_i^{\,}$ are the phonon annihilation operators on site $i$ of the one-dimensional lattice,
$x_i=  b_i^{\dagger} + b_i^{\,}$ is the dimensionless harmonic oscillator coordinate associated
with the atomic vibration on the site $i$, and the sum over $j$ goes over all the nearest-neighbor sites of $i$.
Note that we count energy from the ground state of the local phonon mode.
The minimal physical model for the hopping amplitude dependence on the relative interatomic displacement,
$\Delta R_{ij}$, can be formulated as
\begin{equation}
t(x_i,x_j) =
t_0 e^{ -S (a \mp \Delta R_{ij} )^2/a^2} , \quad \Delta R_{ij} = u (x_i-x_j) ,
\label{txx1}
\end{equation}
with negative sign for hopping in the positive axis direction and positive sign otherwise, see
an illustration in Fig.~\ref{Fig1}. Here $S\gg 1$ is the tunneling action in the rigid lattice with
lattice spacing $a$, and $u=1/\sqrt{2M\Omega}$ is the amplitude of the zero-point motion,
with $M$ and $\Omega$ being, respectively, the mass and frequency of harmonic oscillators associated with the
atomic vibrations. An equivalent re-parametrization can be written as
\begin{equation}
t(x_i,x_j) \equiv  t e^{ \pm g (x_i-x_j) -  \epsilon (x_i-x_j)^2 }\, ,
\label{txx2}
\end{equation}
where $t=t_0 e^{ -S}$, $g=2S(u/a)$, and $\epsilon = S(u/a)^2$. In what follows, we set $t$ as the unit of energy.
The linearized PSSH model corresponds to $t(x_i,x_j) \approx t \pm gt (x_i-x_j)$.

\begin{figure}[tbh]
\includegraphics[width=0.23\textwidth]{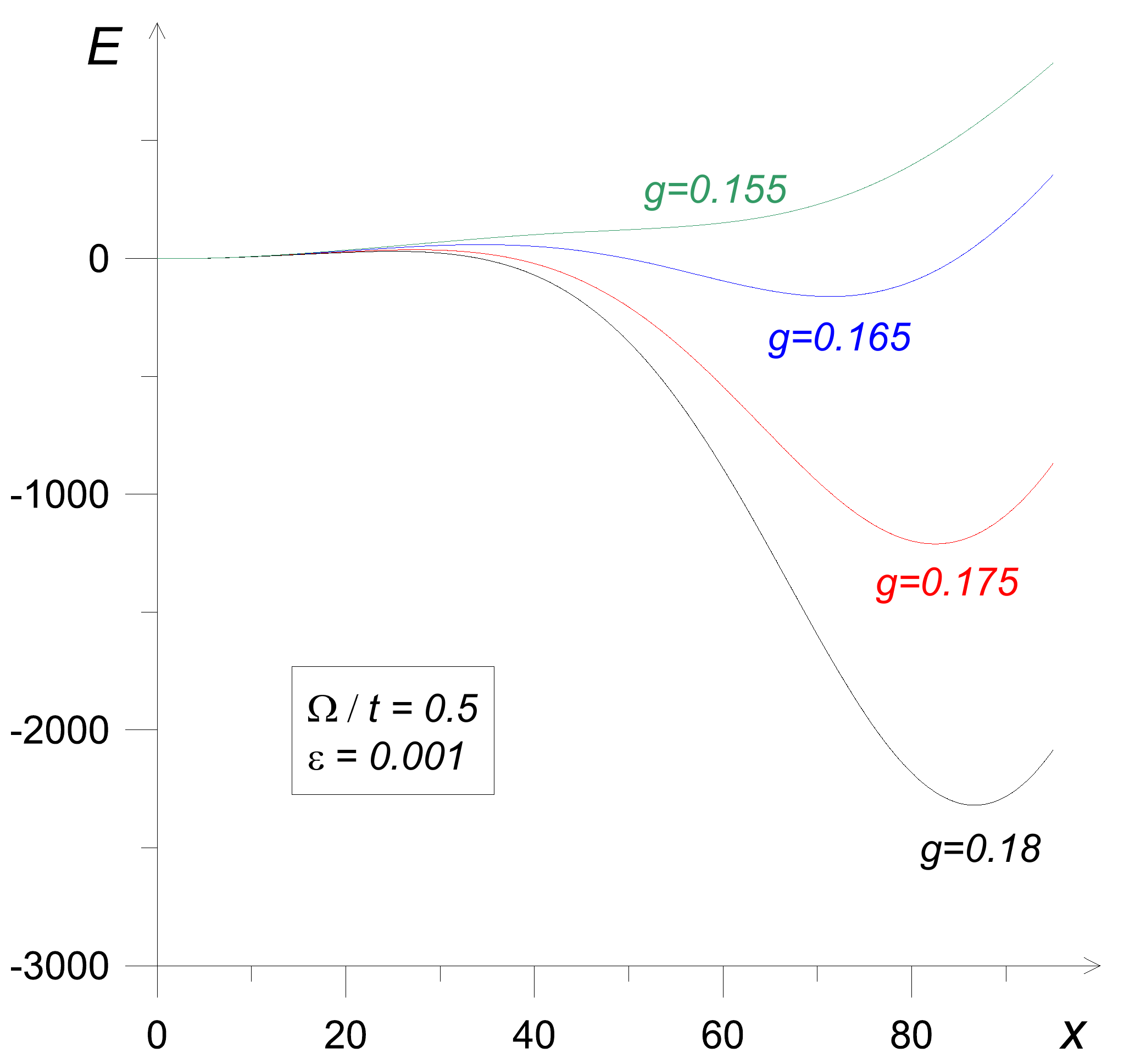}
\includegraphics[width=0.23\textwidth]{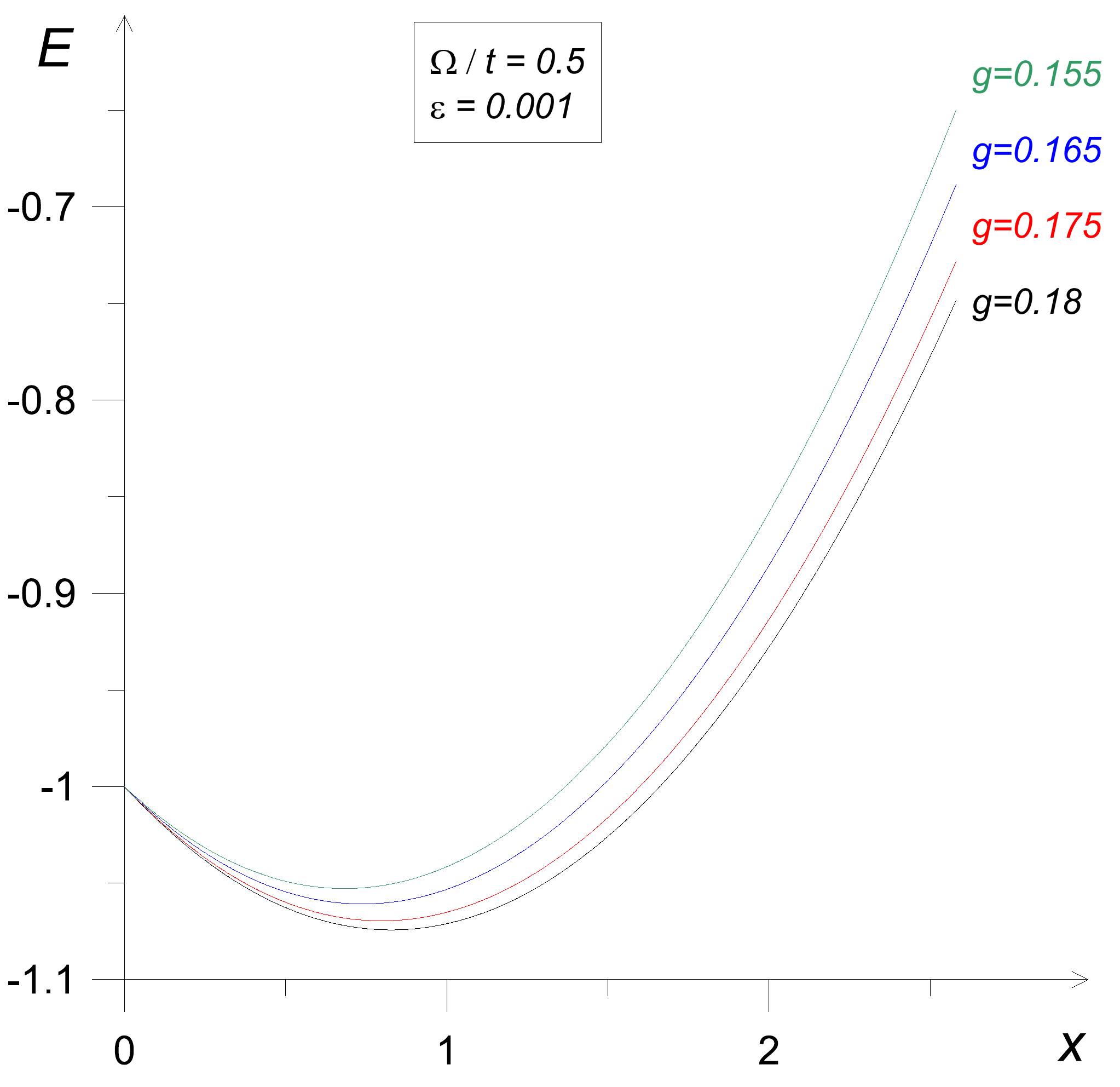}
\caption{Energy landscape (\ref{c1}) for small $\epsilon=0.001$ at large and small values of $x$. }
\label{Fig2}
\end{figure}

To get important insight into model's behavior, consider the energy landscape
\begin{equation}
E(x)= \Omega x^2 /4-t \exp \{ gx-\epsilon x^2\} \;,
\label{c1}
\end{equation}
based on the sum of the harmonic oscillator potential energy and the kinetic-energy gain
by the particle delocalized between the two sites with displacement-dependent hopping
amplitude. At  $\epsilon=0$, the energy is unbounded from below, and the model is pathological.
Finite $\epsilon$ provides the proper regularization. However, as long as $\epsilon$ is very small,
e.g. because of small $u/a$ ratio, the ground state energy, $E_G(g)=\min{E}_x$, undergoes a drastic
crossover at small $g > g_{\epsilon} \sim \sqrt{\epsilon}$ from perturbative expression
$E_G(g) \approx -t -g^2t^2/\Omega$ to exponentially large values
$E_G(g) \propto -t e^{g^2/4\epsilon}$ (see Fig.~\ref{Fig2}).

Qualitatively, the above-described type of behavior is generic for any tight-binding model
characterized by pronounced exponential dependence of hopping on atomic coordinates
with a regularization of $t(x_i,x_j)$ for large displacements
specified by a small parameter---an analog of our $\epsilon$.
All such models feature a sharp self-trapping crossover at a certain $g_{\epsilon} \ll 1$.
This consideration rules out any possibility of having light polarons in model (\ref{H}) and its analogs on approach to
the strong coupling limit when the PSSH electron-phonon coupling is exponentiated $ t(1-gx) \to t e^{-gx}$ and the exponent
is allowed to have large values. It is only in the perturbative regime (i.e., at $g < g_{\epsilon}$)  were the linear and exponential
models of hopping  produce close results.

\begin{figure}[tbh]
\includegraphics[width=0.23\textwidth]{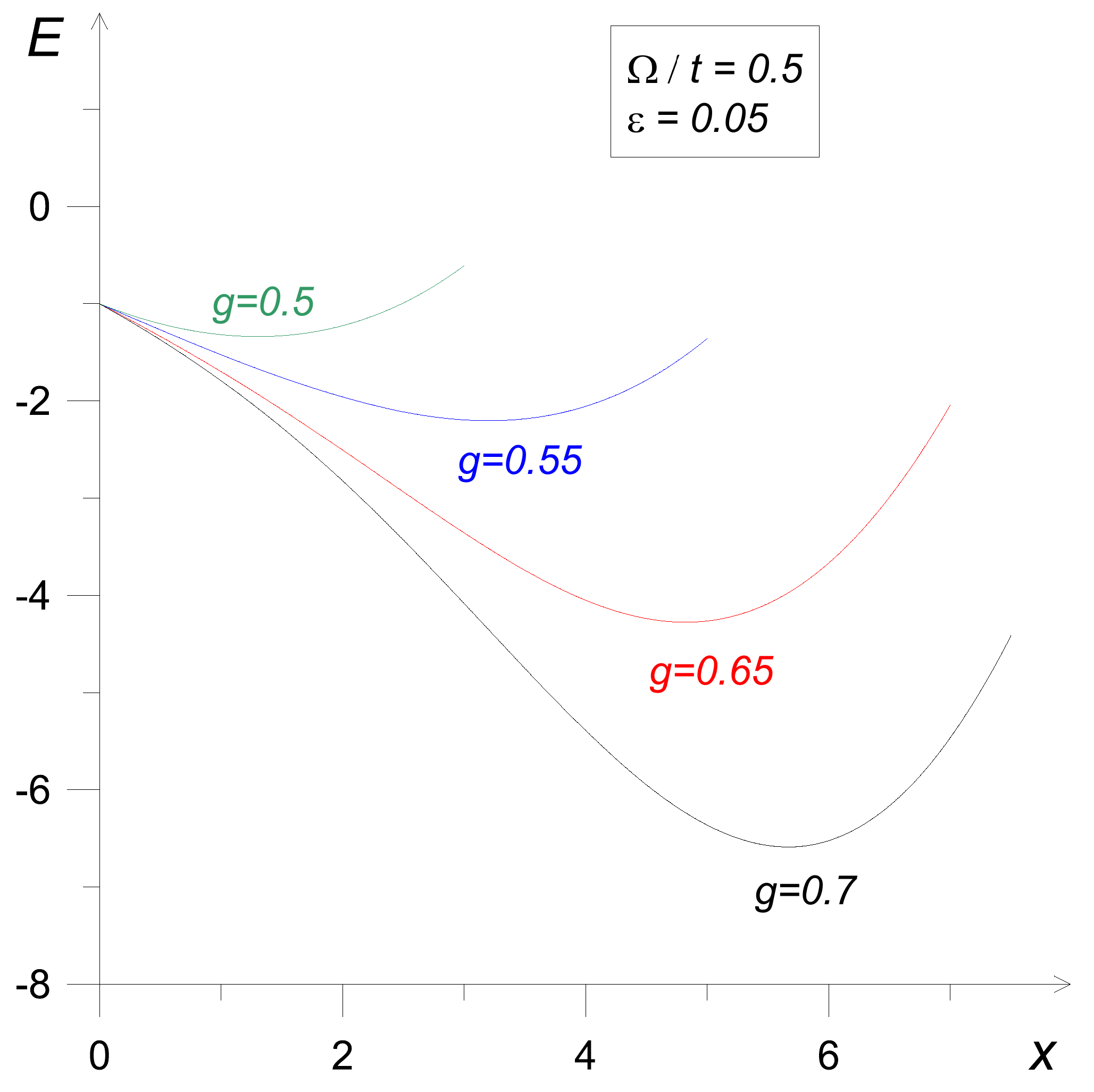}
\includegraphics[width=0.23\textwidth]{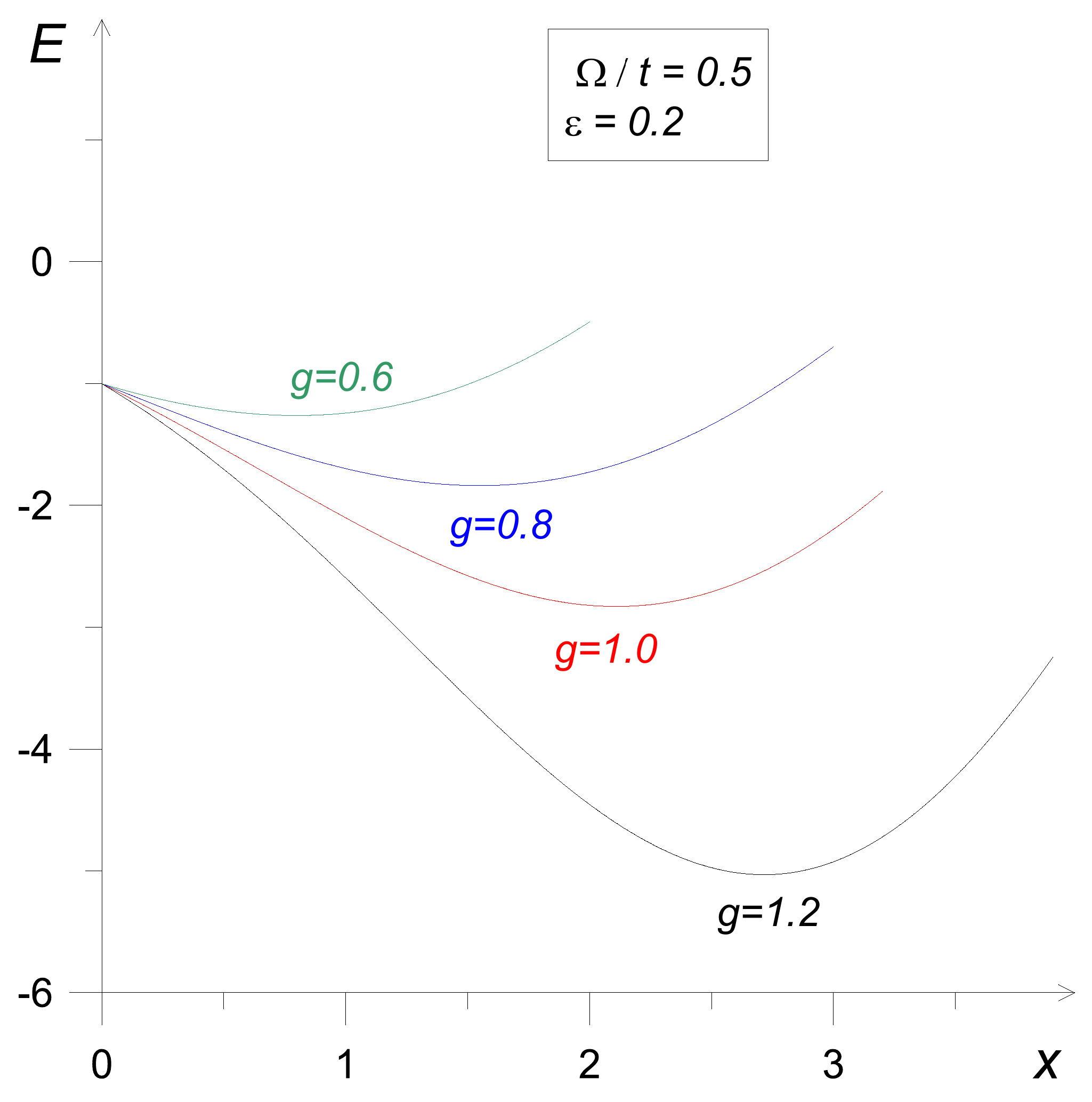}
\caption{Energy landscape (\ref{c1}) at $\epsilon=0.05$ (left panel) and $\epsilon=0.2$ (right panel).}
\label{Fig3}
\end{figure}

When the regularization parameter $\epsilon$ is increased, the situation changes dramatically as illustrated in
Fig.~\ref{Fig3}. Now the energy minimum is unique for any value of $g$. At $\epsilon=0.05$, the crossover to
the exponential energy-gain regime still takes place at moderate values of $g$. At $\epsilon=0.2$, the regime of
strong coupling $g \geq 1$ considered in the previous work can be reached more easily, see right panel in Fig.~\ref{Fig3},
and the most intriguing question is whether the ground state momentum in this case remains zero or shifts to finite
values as in the linearized model.

{\it $X$-representation.}  Within the coordinate representation for both the particle and phonons one
can formulate a sign-positive MC approach to a broad class of polaron problems, with essentially
any kind of particle-phonon interactions---the density-displacement or/and hopping-displacement couplings,
provided the latter are sign-positive. This is readily seen with the standard path integral representation,
where the only condition for the scheme to be sign-positive is the requirement
$t_{ij} \left( \{ x_s \} \right) \geq 0$.

Conventional path-integral treatment of phonons in the $x$-representation would force one to deal with the $(d+1)$-dimensional configuration space, with the finite system size $L$, finite inverse temperature $\beta$, and finite imaginary-time step $\Delta \tau$. Correspondingly, the MC results would need to be extrapolated towards the $L, \beta \to \infty$ and $\Delta \tau \to 0$ limits.
However, harmonic nature of lattice vibrations allows one to partially integrate phonon $x$-paths analytically.
There are two ways of achieving the goal. The first one is particularly suited for dispersionless phonons considered in this work.
An alternative approach utilizes the sign-positive diagrammatic expansion discussed in the Conclusions and Outlook section.

The Gaussian form of the free phonon propagators in the $x$-representation allows one to perform semi-analytic integration over all $x$-variables except for those whose values parametrize the magnitudes of particle hopping amplitudes and thus have to be sampled 
along with the particle worldlines. This naturally leads to the thermodynamic ($L \to \infty$) formulation.
In addition (and along similar lines), the Gaussian form of the ground-state wave function allows one
to formulate an explicit ground-state technique for the imaginary-time polaron Green's function in the
site representation:
\begin{equation}
{\cal G}(\tau, r) = \langle a_r^{\,} (\tau) a_{0}^{\dagger}(0) \rangle \, .
\label{G}
\end{equation}
Here $\tau $  is the imaginary time, $r$ is the (discrete) distance from the origin, 
and $\langle \dots \rangle$ stands for averaging over the ground state of the system.

\begin{figure}[tbh]
\includegraphics[width=0.48\textwidth]{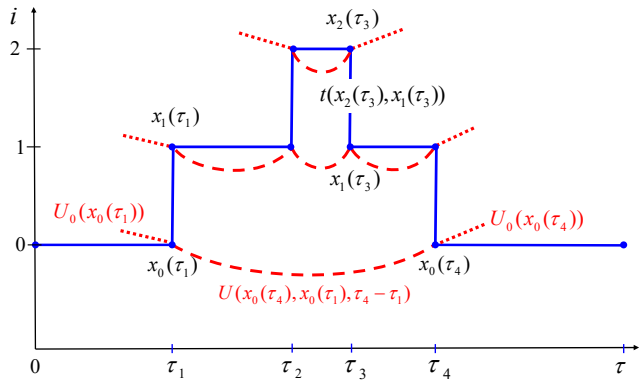}
\caption{Typical $x$-representation diagram for the ground-state Green's function ${\cal G}(\tau,r)$ of the model  (\ref{H}). The blue solid line is the worldline of the particle.  The kinks---the vertical segments of the particle worldline---represent the hopping amplitudes $t$. Dashed and dotted red lines are representing the phonon propagators $U$ and $U_0$, respectively.
Propagators $U_0$ behave as $\tau$-independent vertices because the phonon ground-state energy is set to zero.}
\label{Fig4}
\end{figure}

In general, the necessity of performing macroscopic Gaussian integration when making local updates changing a couple of variables
might bring little advantage compared to sampling full worldline configurations by local updates. However, the gain is dramatic
in the case of dispersionless (i.e., spatially local) phonon modes. Here the paths for phonon modes on different sites are  {\it disconnected}---and thus do not need to be sampled---as long as their $x$-variables are not associated with the hopping events on the particle's worldline.
Projection to the ground state of harmonic oscillators involves special (mixed representation) phonon $x$-propagators
\[
U_0(x, \tau)  = \langle G \vert e^{-\tau H_2 } \vert x \rangle = \langle x \vert e^{-\tau H_2 } \vert G \rangle
\]
connecting the phonon mode ground states $\langle G \vert$ and $ \vert G \rangle$ (at the left and the right ends of the path, respectively) to the corresponding closest in time hopping events controlled by the given mode.
The special propagators $U_0(x, \tau)$ are related to generic phonon $x$-propagators,
\[
U(y,x, \tau ) = \langle y \vert e^{-\tau H_2 }  \vert x \rangle ,
\]
by the obvious relation  $U_0(y,\tau) = \int  U(y,x,\tau ) \, \psi_0(x)\, dx$, where $\psi_0(x)$  is the ground-state wave function of the phonon mode in the $x$-representation. Thanks to our choice of the phonon ground-state energy. $U_0$ propagators are $\tau$-independent
as follows from
\begin{equation}
U_0(y) = \langle y \vert e^{-\tau H_2 }  \vert G \rangle \equiv
\langle y \vert G \rangle = \psi_0(y) =  {   e^{-y^2/4} \over (2\pi )^{1/4}} \, .
\label{U0}
\end{equation}
A typical ``diagram" for ${\cal G}(\tau,r)$ is shown in Fig.~\ref{Fig4}.
The product of all phonon propagators and hopping amplitudes in the graph with $n$
hopping transition constitutes the configuration weight $W_n$. An explicit expression for the propagator
$U(y,x, \tau )$ is given by
\begin{eqnarray}
U(y,x,\tau ) =
 \frac{e^{(1/2) \Omega \tau  - Q(y,x,\tau) } }{  \sqrt{4 \pi \sinh ( \Omega \tau)}  } \, \, , \nonumber\\
 Q(y,x,\tau)\, =\,
\frac{ \cosh ( \Omega \tau) (x^2+y^2) -2xy }{4\sinh ( \Omega \tau ) } \, .
\label{U}
\end{eqnarray}

As long as all the phonon modes are local, adding the density-displacement (e.g., Holstein) couplings on top of the hopping-displacement
ones comes at little computational cost. The effect of standard linear and quadratic in $x$-variables density-displacement interactions 
is readily accounted for by an analytic modification of Gaussian $x$-propagators for every segment of the particle worldline between
two adjacent hopping events (linear and quadratic density-displacement couplings lead to a modified shifted harmonic oscillator Hamiltonian); the cost for treating generic density-displacement interactions is also moderate--—numeric tabulation
of the $x$-propagator. If the density-displacement coupling is to the very same modes that control the value of the hopping amplitude, then we deal with exactly the same diagrams as in Fig.~\ref{Fig4}, but now with modified $x$-propagators. If the density-displacement coupling is to separate phonon modes, one needs to introduce propagators for those modes and sample
the corresponding $x$-variables specified at all sites connected by the particle hopping transitions.

{\it Monte Carlo scheme.} The sign-positive diagrammatic-type expansion for $G(\tau, r)$ in powers of hopping transitions
leads to a simple and efficient diagrammatic Monte Carlo \cite{PST1998,PS1998} simulation protocol.
Our scheme is based on updates that change (i) the variable $\tau$ [the $\tau$-update], (ii) the variable
$x_{r_{k}}(\tau_k)$ or $x_{r_{k}}(\tau_{k-1})$ [the $x$-update], and (iii) the expansion order, $n$, by $\pm 1$ [the $(n\pm 1)$-updates].

$\bullet$ In the $\tau$-update, the new value $\tau ' > \tau_n$ is proposed from the exponential probability distribution
$P(\tau ') = - e^{\mu (\tau' - \tau_n)}/\mu$. Here $\tau_n$ is the time moment of the $n$-th  (i.e., the last one in time domain)
kink and $\mu < 0 $ is an auxiliary parameter introduced for controlling the Green's function statistics in the time domain.
This update is always accepted. (We do not mention here standard for all MC simulations tools for generating random
variables from arbitrary probability distributions and the flat-histogram sampling).

$\bullet$ In the $x$-update, we select at random one of the hopping transitions and propose to update the oscillator
coordinate at one of the two sites involved. This update

\begin{figure*}
\subfigure{\includegraphics[width=0.3\textwidth]{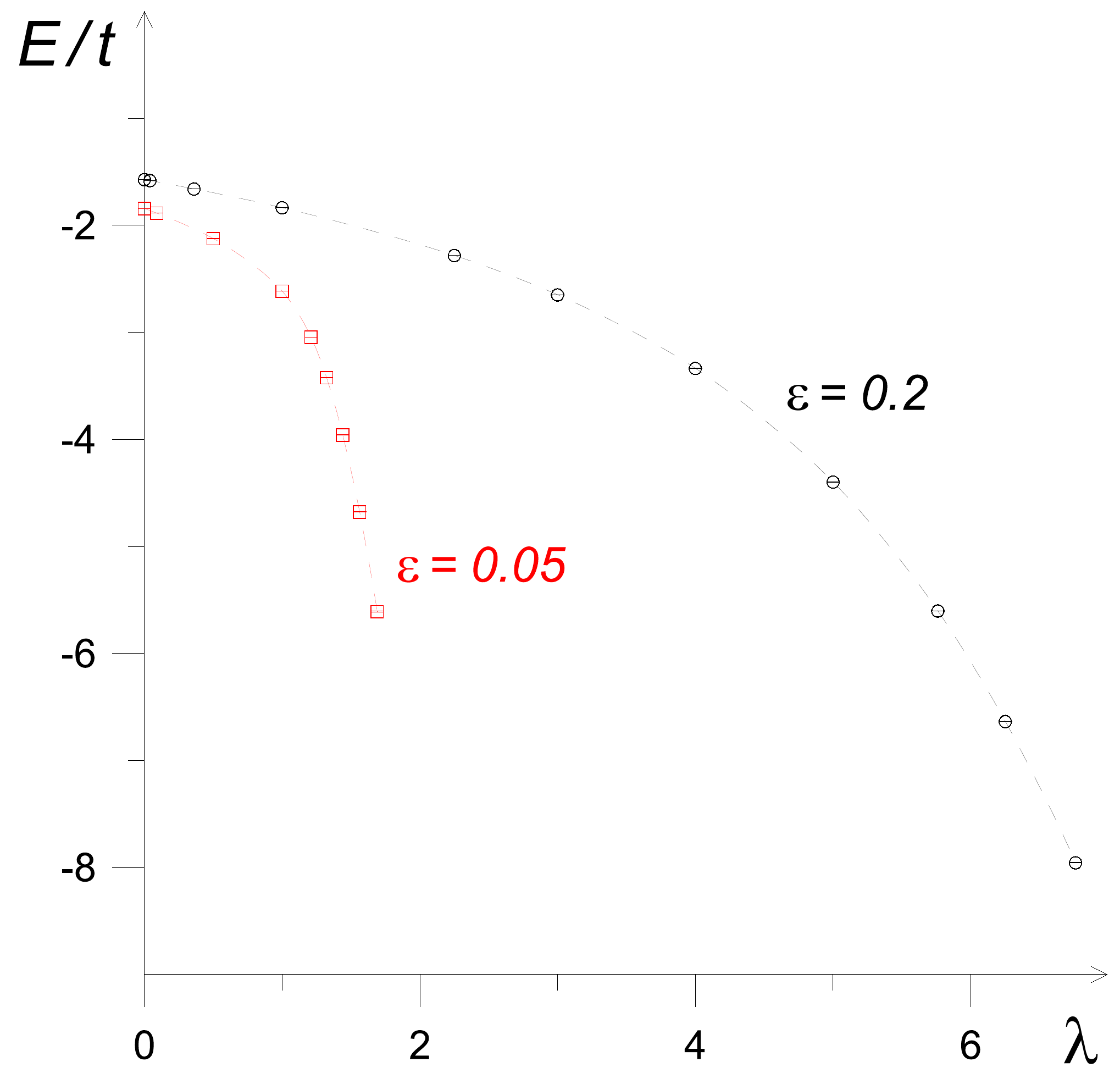}}
\subfigure{\includegraphics[width=0.3\textwidth]{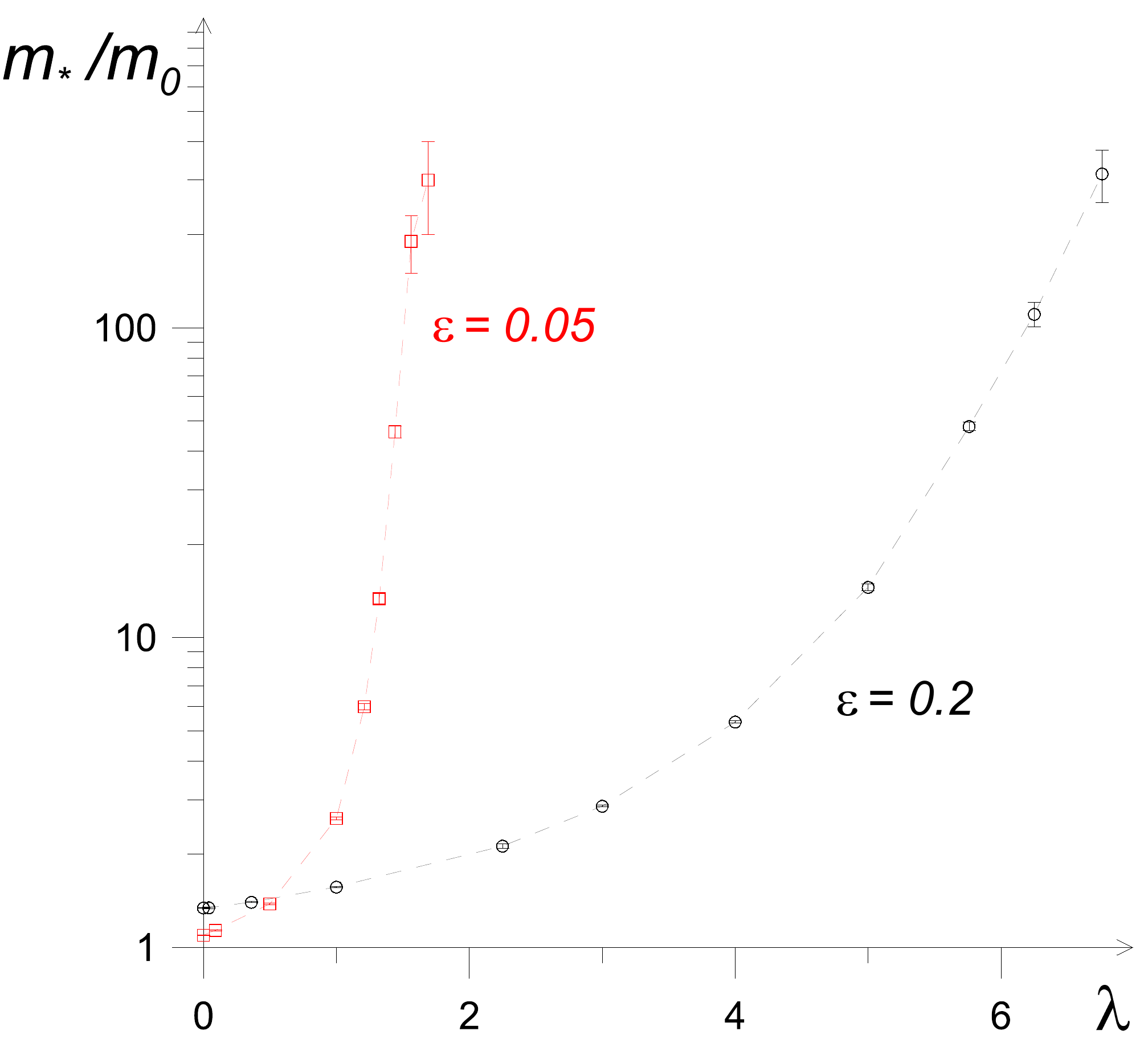}}
\subfigure{\includegraphics[width=0.3\textwidth]{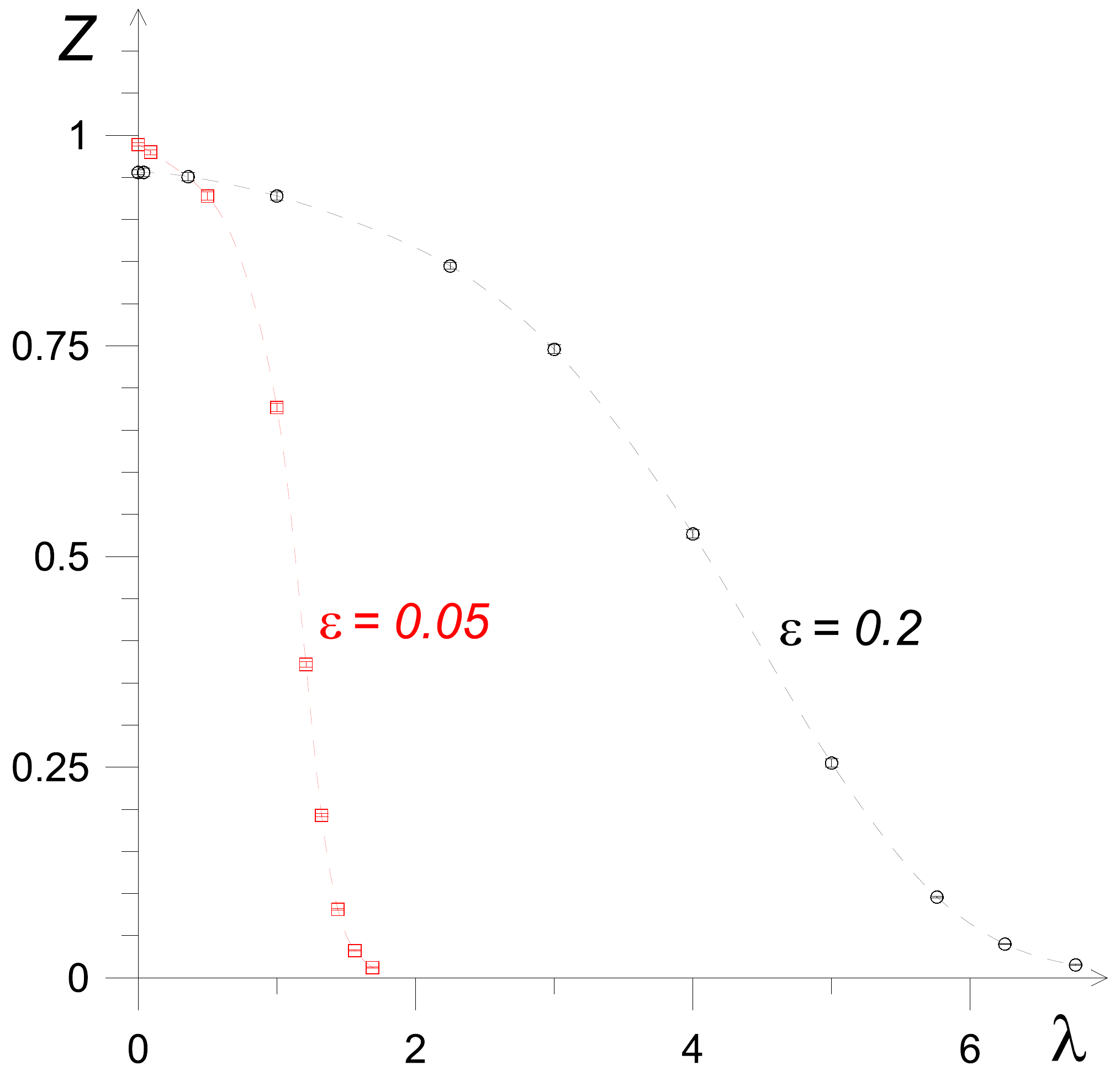}}
\caption{Energies, effective masses, and quasiparticle residues as functions of $\lambda=2g^2t/\Omega$ for two values
of regularization parameters $\epsilon=0.2$ (black dashed curves with open circles) and $\epsilon=0.05$
(red dashed curves with open squares). Error bars are shown in all plots; for energy values they are orders of magnitude smaller than symbol sizes.}
\label{Fig5}
\end{figure*}
\FloatBarrier \noindent
changes the product of all
phonon propagators depending on this coordinate times the value of the hopping amplitude. Since
all the functions involved are Gaussian functions of the updated oscillator coordinate,
the $x$-update is rendered rejection-free by proposing the new coordinate, $x'$, from the Gaussian distribution
\[
P(x') = (2\pi \sigma^2 )^{-1/2} \exp [ -(x'-z)^2/2\sigma^2 ]
\]
with the shift $z$ and dispersion $\sigma$ depending on other relevant graph variables.

$\bullet$ In the $(n+1)$-update, called with probability $p_+$, we propose to insert a hopping event at
time $\tau_{n+1} $ selected from the uniform distribution on the interval $(\tau_n,\tau )$. The direction of the
transition from site $R=r_n(\tau_{n})$ is selected at random; this defines lattice site $R'=r_{n+1}(\tau_{n+1})$.
The new oscillator variables $x_R(\tau_{n+1})=y$ and $x_{R'}(\tau_{n+1})=z$ are proposed from the Gaussian distribution
\[
P(x,y) \propto e^{-(y^2+z^2)/2 \pm g(y-z) -\epsilon (y-z)^2}.
\]
In the $(n-1)$-update, called with probability $p_-$,
the last hopping event (if there is one; otherwise the update is rejected) is simply erased from the configuration. The acceptance
ratio for the complementary pair of the  $(n\pm 1)$-updates  equals
\begin{equation}
R \, =\,  \frac{p_-}{p_+} \frac{W_{n+1}}{W_{n}} \frac{2  (\tau - \tau_{n}) \, t(y,z)}{P(y,z)} \; .
\label{R}
\end{equation}

{\it Results.} The full quantum mechanical solution of the problem confirms the overall picture established
on the basis of the energy landscape (\ref{c1}) and finds that the dispersion minimum remains at zero momentum
when the polaron $Z$-factor collapses to near zero values, the effective mass undergoes an explosive enhancement,
and the ground-state energy exceeds the bare particle half-bandwidth, see Fig.~\ref{Fig5}.
All these effects are typical for transition to the nearly localized state at strong coupling.

{\it Conclusions and Outlook.}  Our results show that unusual properties characteristic of linearized
PSSH polarons may be absent in models with sign-definite $t_{ij}$ thereby emphasizing importance of non-linear terms
at strong coupling in physical systems, and challenging strong-coupling predictions obtained for models with linearized
sign-alternating hopping $t_{ij}$. In this respect, special attention should be paid to the
microscopic physics leading to sign-alternating phonon-modulated hopping, such as, e.g., 
phonon-controlled competition between the tunneling paths in multi-orbital systems \cite{Zhang2022b}.

Our results were obtained with the MC approach based on the diagrammatic technique combining the worldline representation for the particle
with the $x$-representation for atomic displacements. The method applies to a broad class of polaron problems with density-displacement and hopping-displacement couplings, provided the former are harmonic and the latter are sign-positive. A dramatic simplification
of the scheme, with the associated efficiency gain, takes place in models with dispersionless phonos.

\begin{figure}[tb]
\includegraphics[width=0.48\textwidth]{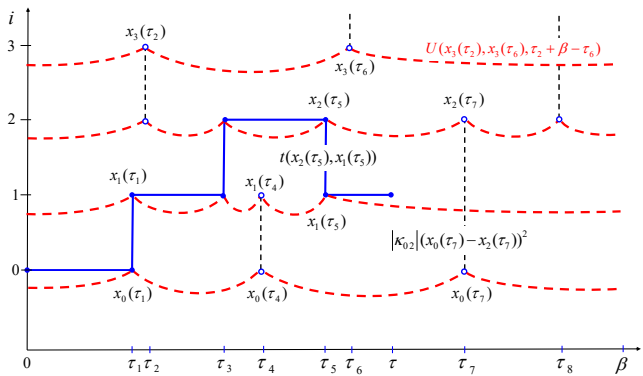}
\caption{Typical $x$-representation diagram for the Green's function ${\cal G}(\tau,r)$ for the dispersive-phonon counterpart of the model  (\ref{H}).  The particle worldline and the phonon propagators in the $x$-representation have the same meaning as in Fig.~\ref{Fig4}.
The new elements are the vertical dashed lines representing the attractive interaction between auxiliary local modes (cf. \cite{Boninsegni2006}). Note also that now the diagram is extensive: It occupies a macroscopic space-time volume.}
\label{Fig6}
\end{figure}

In the case of dispersive phonons, the most challenging aspect of exact formulation
is the necessity of performing an extensive (macroscopic) integration of all atomic coordinates.
There is, however, an option of introducing a scheme where all Monte Carlo updates remain local in space-time
and, thus, computationally efficient. Here we observe that dispersive
phonon modes can be always represented as the result of \textit{attractive} interactions between local modes.
The standard diagrammatic expansion in powers of the attractive interaction potential in the
$x$-representation is then sign-positive and allows efficient MC sampling (cf. \cite{Burovski2006,Boninsegni2006}); see Fig.~\ref{Fig6}. The price we pay for having local/intensive updates is the extensive
character of the configuration space and the necessity of extrapolating to the
thermodynamic ($L\to \infty$) and ground-state ($\beta \to \infty$) limits.
An explicit decomposition of the phonon potential energy into fictitious local part and attractive
interactions is as follows (note that $\kappa_{ij} = \kappa_{ji}$, $\kappa_{ii}>0$):
\[
\sum_{ij} \, \kappa_{ij} \, x_i x_j \, =\, \sum_i \, \tilde{\kappa}_i x_i^2 - \sum_{i<j} |\kappa_{ij} | [x_i - {\rm sgn} (\kappa_{ij})  \, x_j  ]^2 \, ,
\]
\[
\tilde{\kappa}_i\, =\, \kappa_{ii}\, +\, \sum_{j\neq i}  |\kappa_{ij}| .
\]

{\it Acknowledgments.} We acknowledge inspiring discussions with Andrew Millis, David Reichman, Mona Berciu, Chao Zhang, and John Sous.
This work was supported by the National Science Foundation under Grant No. DMR-2032077.

\end{document}